\documentclass[a4paper,11pt]{article}
\pdfoutput=1 
\usepackage{jinstpub} 
\usepackage{natbib}
\usepackage{dcolumn}
\usepackage{upgreek}
\usepackage[dvipsnames]{xcolor}
\usepackage{graphicx}
\usepackage{hyperref}
\usepackage[load-configurations=abbreviations,detect-all=true]{siunitx}
\usepackage[version=3]{mhchem}
\usepackage[mathlines]{lineno}

\usepackage{xspace}

\newcommand{\bestppm}{$<\SI{10}{ppm}$\xspace}

\newcommand{\TMIIm}{\mbox{\emph{Topmetal-II\raise0.5ex\hbox{-}}}\xspace}

\newcommand{\MOLLER}{MOLLER\xspace}
\newcommand{\Moller}{M{\o}ller\xspace}

\DeclareSIUnit\keV{keV}
\DeclareSIQualifier\ee{ee}
\DeclareSIQualifier\nr{nr}
\DeclareSIUnit\pe{p.e.}

\usepackage{array}
\newcolumntype{L}[1]{>{\raggedright\let\newline\\\arraybackslash\hspace{0pt}}m{#1}}
\newcolumntype{C}[1]{>{\centering\let\newline\\\arraybackslash\hspace{0pt}}m{#1}}
\newcolumntype{R}[1]{>{\raggedleft\let\newline\\\arraybackslash\hspace{0pt}}m{#1}}

\title{\boldmath A direct-sampling RF receiver for MOLLER beam charge measurement}

\author[a,1]{Yuan Mei\note{Corresponding author.},}
\author[a]{Shujie Li,}
\author[a]{John Arrington,}
\author[a,2]{Joseph Camilleri\note{Currently at Virginia Tech, Blacksburg, VA},}
\author[b]{Aled Cuda,}
\author[b]{James Egelhoff,}
\author[a,b]{Yury Kolomensky,}
\author[a]{and Ernst Sichtermann}

\affiliation[a]{Lawrence Berkeley National Laboratory, Berkeley, California, USA}
\affiliation[b]{Department of Physics, University of California, Berkeley, USA}

\emailAdd{ymei@lbl.gov}

\abstract{We have developed and tested a direct-sampling RF receiver capable of measuring the amplitude of a \SI{1497}{MHz} sinusoidal signal in \SI{0.5}{ms} integration windows to within \bestppm relative uncertainty.  The receiver is intended for measuring signals from beam current monitoring cavities on the beamline of the Continuous Electron Beam Accelerator Facility (CEBAF) at Jefferson Laboratory.  The signal strength, frequency, and integration window are consistent with the thus far unmet requirements of the upcoming MOLLER experiment to measure the beam charge for different helicity states.}

\begin{document}
\maketitle
\flushbottom

\section{Introduction}
\label{sec:intro}

Precision measurements of parity violation in electron scattering are a powerful probe of New Physics, complementary to direct searches at high-energy colliders.  The \MOLLER experiment\cite{Benesch:2014bas} has been proposed to carry out an ultra-precise measurement of the parity-violating left-right asymmetry in the \Moller scattering of longitudinally polarized \SI{11}{GeV} electrons produced at the Continuous Electron Beam Accelerator Facility (CEBAF) off electrons in a liquid hydrogen target in Hall A at Jefferson Laboratory.

A successful measurement would provide unprecedented sensitivity to physics beyond the Standard Model, particularly purely leptonic contact interactions beyond the reach of existing high-energy colliders, as well as to \SI{100}{MeV}-scale dark photons that might have small mixing with the $Z^0$ boson.  It would achieve an unprecedented determination of the weak mixing angle at low energy.

\MOLLER aims to measure the parity-violating left-right asymmetry $A_{PV}\equiv(\sigma_R-\sigma_L)/(\sigma_R+\sigma_L)$ of 35 parts per billion (\si{ppb}) with a precision of \SI{0.8}{ppb} (\SI{2.4}{\percent}).  The asymmetry in the scattering cross-section $\sigma_{L(R)}$ is measured by changing the helicity of the incident beam from left (L) to right (R) in \SI{0.5}{ms} intervals.  Achieving this precision requires an unprecedented control and measurement of the parameters of the incident electron beam: beam charge, first moments of the beam distributions (position, angle, energy), and the second moments of the beam distributions (beam size).  Helicity-correlated variations of the beam parameters can induce false asymmetries comparable to or larger than the physics asymmetry and must thus be controlled and measured in real time.  \MOLLER requires the beam charge to be measured with the relative precision of 10 parts per million ($10^{-5}$) per \SI{0.5}{ms} window and beam size with the relative precision of $10^{-4}$.

The standard solution for measuring the position and intensity of CEBAF beam is to pick up and digitize a beam-induced signal from a microwave cavity\cite{Farkas:1976slac,Whittum:1998uc}.  The main challenge for \MOLLER is the linearity and dynamic range of the microwave receivers used to process the signal.  Modern systems utilize single or multi-stage heterodyne detection of the RF amplitudes and phases with mixing to intermediate frequencies (IF) in \si{MHz} range, followed by digitization in commercial waveform digitizers and online or offline data processing.  This system has obvious limitations.  The linearity and the dynamic range of a heterodyne receiver are limited by the RF mixer(s). The signal-to-noise ratio has the ultimate floor determined by the quality of the local oscillator (LO).  Achieving noise levels of \SI{10}{ppm} (\SI{100}{dB} SNR) in a \SI{2}{kHz} bandwidth around \SIrange{1}{2}{GHz} LO frequency is very challenging.  The best resolution of the beam charge measurement at Jefferson Laboratory was achieved by the $Q_\mathrm{weak}$ experiment, where their state-of-the-art heterodyne receivers  showed the noise floor of about \SI{40}{ppm} per \SI{960}{Hz} helicity window\cite{Qweak:2014xey}, dominated by the LO noise.

We aim to develop and implement a novel fully digital microwave processor based on a fast, \SI{3}{Gsps} 14-bit digitizer.  Recent advances in digital electronics, both in fast digitizers and onboard FPGA signal processing make direct, digital signal processing (DSP) at the carrier frequency of \SI{1.5}{GHz} possible.  A photograph of the fully digital receiver prototype is shown in Fig.~\ref{fig:ReceiverTopView}.  The receiver digitizes the cavity signal directly (after appropriate amplification and bandpass filtering), applies a digital filter to the digitized waveform, performs digital down-conversion (DDC) of the waveform, measures the amplitude and phase of the signal for each helicity window, converts them to charge, and sends the output data to the data acquisition system.  This scheme has a significant advantage in terms of the ultimate noise floor.  LO amplitude noise does not contribute to the resolution of the processor, and LO phase noise can be very small.  We estimate that a 10-bit (effective) commercial digitizer sampling at \SIrange{3}{4}{Gsps} has an ultimate noise floor of \SI{1}{ppm}, more than sufficient for the needs of the next-generation parity experiment.  Even with a reasonable noise figure of RF amplifiers and allowing for some other nonidealities, the resolution requirement of \SI{10}{ppm} should be within reach.

In this paper, we show that using commercial off-the-shelf parts, we successfully constructed an RF receiver that reached the $<\SI{10}{ppm}$ resolution \MOLLER requires.  It provides a tool for diagnostics at this uncertainty level, which is currently unattainable by other means.  Observation of additional noise at \SI{25}{ppm} level from Hall-A beam using this receiver is also discussed.

\section{RF receiver design}

The primary function of the RF receiver is to measure the RF power in fixed integration windows with the best signal-to-noise ratio possible.  The nominal input signal is sinusoidal at \SI{1497}{MHz}, which is the 3rd harmonic of the beam frequency \SI{499}{MHz}, and the target integration window is about \SI{0.5}{ms}.

\begin{figure}[!htb]
  \centering
  \begin{minipage}[c]{0.5\linewidth}
    \includegraphics[width=\linewidth]{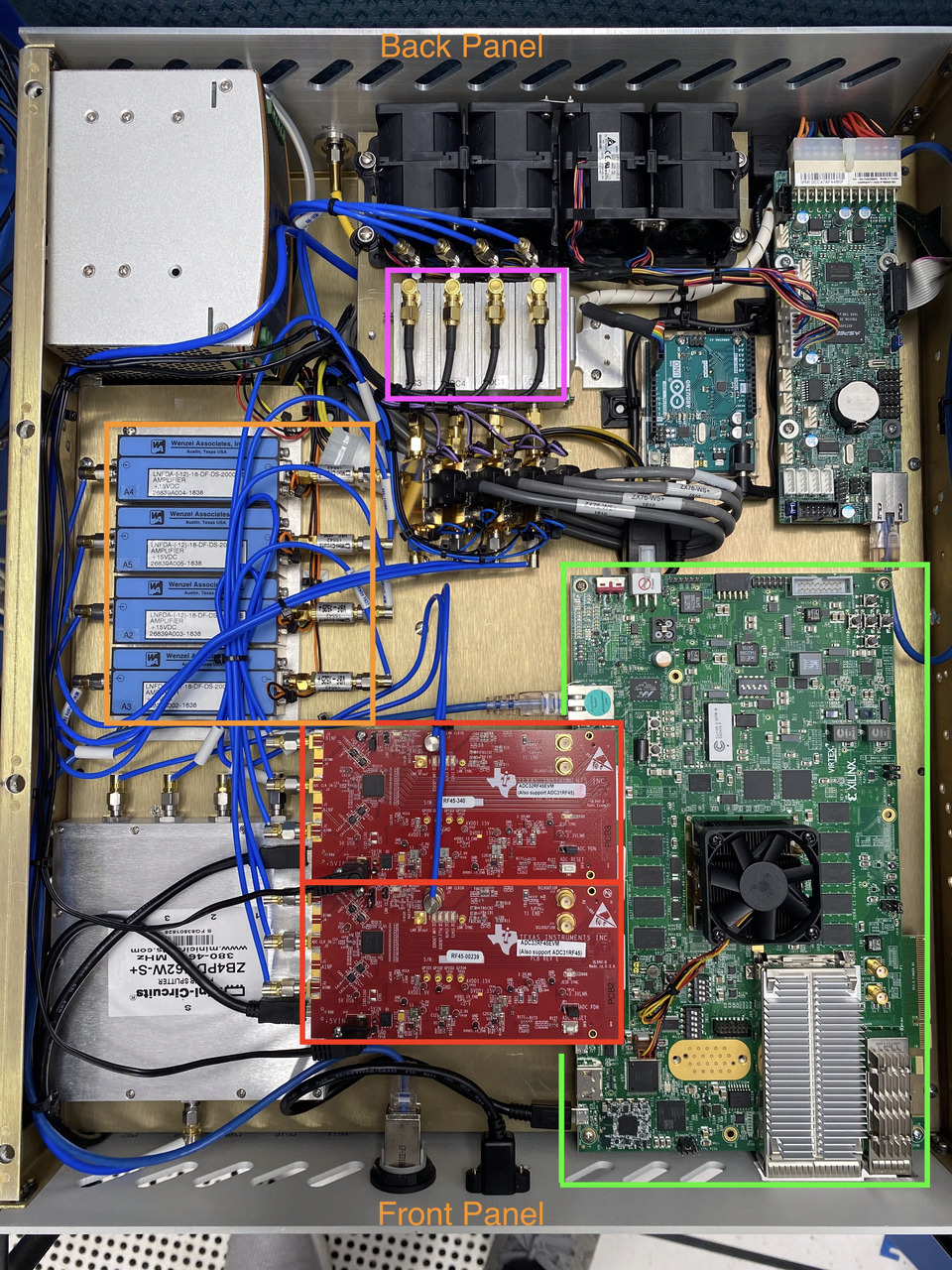}
  \end{minipage}\hspace{2em}%
  \begin{minipage}[c]{0.2\linewidth}
    {\setlength{\fboxrule}{2bp}
     \fcolorbox{green}{white}{VCU108 FPGA board}\\
     \fcolorbox{red}{white}{ADC32RF45EVM}\\
     \fcolorbox{orange}{white}{Amplifier and filter}\\
     \fcolorbox{pink}{white}{Directional coupler}
    }\\
  \end{minipage}
  \caption{Top view of the receiver.  The chassis is a standard 2U, 19-inch rack enclosure.  The power and fan-based cooling are controlled by an IPMI-enabled controller shown in the top-right.  Four RF and one clock input connectors are mounted on the back panel.  The clock is split 4 ways via a power splitter in the bottom-left to drive the two ADCs.
  }
  \label{fig:ReceiverTopView}
\end{figure}

The receiver consists of three main components: an RF front-end that brings the signal to an appropriate amplitude and preliminarily rejects out-of-band noise and interference, a high-speed ADC that directly digitizes the RF signal, and an FPGA that processes the data and transmits the data to a computing server.  A photograph of the completed receiver is shown in Fig.~\ref{fig:ReceiverTopView}.

The core of the system is a commercially available high-speed ADC (ADC32RF45, \emph{Texas Instruments, Dallas, TX}).  This ADC is capable of a sampling rate of \SI{3}{Gsps} and an analog input bandwidth up to \SI{4}{GHz}\cite{adc32rf45ds:2020}.  It can place the input signal of \SI{1497}{MHz} in the 1st Nyquist zone.  ADC32RF45 has a noise spectral density of \SI{-155}{dBFS/Hz} and 9.5 effective number of bits (ENOB) for \SI{1}{GHz} input signal at a sampling rate of \SI{3}{Gsps}.  Besides the sampling rate and noise performance, another importance reason for choosing this ADC is the availability of an evaluation module (EVM) for prototyping and its high-speed serial link to FPGA.  The rest of the system is designed around this ADC.


\subsection{RF front-end}

\begin{figure}[!htb]
  \centering
  \begin{minipage}[t]{0.9\linewidth}
    \includegraphics[width=\linewidth]{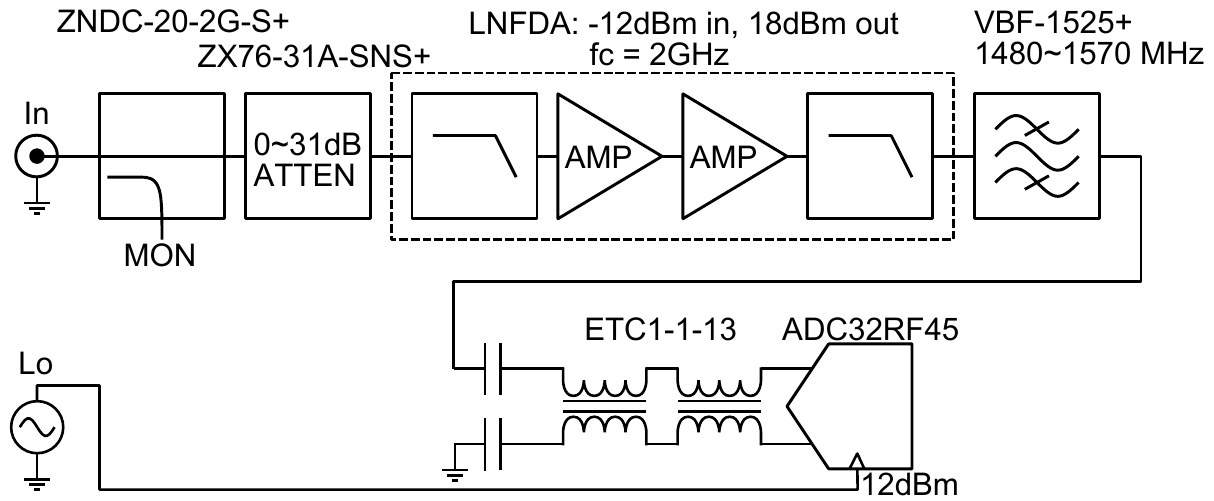}
  \end{minipage}\hspace{0em}%
  \caption{The RF chain of the sampling system with their part numbers.}
  \label{fig:RFchain}
\end{figure}

The RF signal from a standard TM$_{010}$ mode cavity is conditioned through a chain that consists of an attenuator, an amplifier, and a bandpass filter before entering the ADC.  As shown in Fig.~\ref{fig:RFchain}, the input signal first goes through a directional coupler.  Its ``coupled'' output, \SI{-20}{dB} of the main signal, is reserved for monitoring purposes by an external diagnostic instrument.  The signal then passes through a programmable step attenuator.  This attenuator is the sole means for controlling the magnitude of the signal seen by the ADC.  The signal is then fed into a filtered dual amplifier (LNFDA by Wenzel Associates, \emph{Austin, TX}) before passing through a bandpass filter.  The gain provided by the amplifier is necessary to compensate for the large signal power loss in the cable between the cavity and the receiver.  The bandpass filter's center frequency and bandwidth are selected to loosely enclose the expected input signal of \SI{1497}{MHz}.  Finally, the conditioned signal, which is single-ended, is converted to differential and coupled to the ADC via a capacitor-transmission line transformer circuit.  The single-ended to differential conversion circuit is part of the ADC's evaluation module (EVM).  All other components for RF signal conditioning are commercial off-the-shelf and connected together using standard SMA coax cables.  The part numbers are marked in Fig.~\ref{fig:RFchain}.

There are 4 identical RF front-end chains implemented in this system.

\subsection{Digitization}

We chose the ADC manufacturer's evaluation module (EVM, part number: ADC32RF45EVM) for the convenience of prototyping.  The EVM plugs into the FPGA module directly via a industry-standard high-pin-count (HPC) FMC connector.  The sampling clock and RF inputs are standard SMA connectors.  The signal inputs of the EVM are single-ended, which match the outputs of the RF front-end.  Proper coupling to the ADC chip is achieved on the EVM without user intervention.  The digitized signal by the ADC, in the form of high-speed serialized data, is transmitted into an FPGA via an FMC connector.  The data link protocol between ADC and FPGA is JESD204.  The ADC configuration is done using the vendor-provided software via a USB connection to the evaluation board.  Each EVM has two independent ADC channels.  The system incorporates 2 EVMs to implement a 4-channel receiver (Fig.~\ref{fig:ReceiverTopView}).

We chose the Virtex UltraScale FPGA VCU108 Evaluation Kit (part number: EK-U1-VCU108-G, by Xilinx, \emph{San Jose, CA})\cite{vcu108:2020} as the main processing unit of the system.  VCU108 is one of the very few commercially available modules capable of handling two ADC32RF45EVM outputs with appropriate (HPC) FMC interfaces and immense data rates.  The maximum aggregate data rate of two ADC32RF45EVM, which is 4 ADC channels combined, exceeds \SI{144}{Gbps}.

The sampling clock is provided by an ERASynth RF signal generator\cite{erasynth:2020}.  The subsequent digital data clock is derived from the sampling clock.  The processing in the FPGA is synchronous to the data clock.  The clock generator was chosen mainly due to its excellent phase noise, specified at 10, 100, \SI{1}{k}, and \SI{10}{kHz} as -66.1, -100.7, -117.1, and \SI{-126.2}{dBc/Hz} respectively.  The broadband phase noise is at \SI{-154.3}{dBc/Hz}.  The clock generator also has a \SI{10}{MHz} reference interface.  It provides an opportunity to fully phase-lock the sampling to the accelerator beam, which derives its clock from the same \SI{10}{MHz} reference.  The exact sampling rate is set to \SI{3072}{Msps}.  This particular frequency is one of the nominal frequencies widely used in telecommunication.  By adhering to the nominal frequencies, we reduce the complexity in configuring transceivers and IP blocks in the FPGA.

The ADC could be configured to stream raw samples that are truncated to \SI{12}{bits} per channel at above \SI{2.5}{Gsps}.  It could also perform an internal digital down-conversion (DDC) by mixing the digitized raw signal with a numerically-controlled oscillator (NCO) signal.  In essence, the digital circuits compute $\sin$ and $\cos$ values and multiply the them with the digitized signal, then stream the in-phase ($I$) and quadrature ($Q$) numbers.  The ADC also applies a loose Finite Impulse Response (FIR) digital filter then decimates the stream by a selectable factor of 4 to 32, which greatly reduces the data rate.  In the DDC/decimation mode, $I$ and $Q$ numbers are 16-bit each.  The FPGA can receive data of either mode continuously, further reduce the data, then measure the signal amplitude in each time window of interest.

To measure the signal amplitude in each helicity window precisely, as required by the \MOLLER experiment, the start and the duration of the measurement (integration) window must be precisely aligned with the signal that drives the helicity flips.  The FPGA accepts the helicity flip signal from a digital channel that is separate from the ADC.  In this way, both the raw ADC samples and the helicity flip signals are available in the same clock domain inside the FPGA.  The integration window is then defined by synchronizing to the helicity flip and adding appropriate phase shifts.  The phase delay and the window size are then fully programmable.

We transmit the data from FPGA to a server via a \SI{100}{gigabit} ethernet link using raw packets.  On the server, we built a requisition software on top of the DPDK\cite{dpdk} package to directly grab packets from Network Interface Card (NIC) to memory to achieve near-line-rate data reception speed.  We generally transmit all raw data to the server for prototyping so that the processing algorithms could be implemented in software instead of the firmware.  This strategy greatly reduces the development time for validating the processing algorithms.  For future deployment of the system in the experiment, we shall implement proven algorithms in the FPGA to reduce the data to a form and rate appropriate for the data acquisition system employed by experiment. 

\subsection{Data processing}
\label{sec:dp}

We approach the signal amplitude measurement in two ways.  One is to compute the root-mean-square (RMS) in each integration window; the other is to first down-convert the signal to DC (\SI{0}{Hz}), then calculate the average in each window.

Computing the RMS is straightforward.  There are two potential caveats.  The first is that
the ADC output may have a DC offset, and RMS will capture it.  Our tests show that the DC offset is negligible for all integration window sizes of interest.  Also, since the application is only interested in the relative difference between consecutive windows, the DC offset, if exists, will cancel out.  Therefore, we do not explicitly remove DC in the algorithm.  The second is that each integration window must enclose the same fraction of a sinusoidal signal beyond an integer number of periods.  If the remainder contributes to each window differently, depending on the phase, it will add additional fluctuations to the computed RMS.  We circumvent this issue by phase-locking the input signal with the sampling clock and selecting the integration window size to be exactly an integer number of signal periods.

Computing RMS to estimate amplitude is valid for both direct ADC samples and for the $I$ and $Q$ streams after DDC/decimation.

For down-converting the signal to DC, we also ensure that the input signal is phase-locked to the sampling clock and the input signal frequency is a rational factor of the sampling frequency.  Under these conditions, the incoming sinusoidal signal, sampled by the ADC, can be expressed as
\begin{equation}
    x_i = A\cos(2\pi\frac{p_0}{q_0}i)\,,
\end{equation}
where $x_i$ denotes the signal sampled at time point $t_i$, $p_0/q_0$ is a rational number and $i$ is an integer index that labels the sequence of samples.  The input signal's frequency $f_0$ is related to the sampling frequency $f_s$ by $f_0=\frac{p_0}{q_0}f_s$.

When the DDC/decimation mode in ADC is used, $x_i$ is not transmitted to the FPGA.  Instead, the in-phase and quadrature components $I$ and $Q$ are numerically computed from $x_i$ using the relations
\begin{align}
 I_i &= x_i\cos(2\pi\frac{p_1}{q_1}i+\phi)\\
 Q_i &= x_i\sin(2\pi\frac{p_1}{q_1}i+\phi)\,.
\end{align}
The rational number $p_1/q_1$ defines an NCO (digital ``LO'') of frequency $f_n$ that is related to the sampling frequency by $f_n=\frac{p_1}{q_1}f_s$.  This step effectively implements an I/Q mixer that mixes the input signal with an LO of frequency $f_n$ in digital domain inside of the ADC.  The phase $\phi$ is fixed upon the reset of ADC and remains constant afterward but changes to a different value after each reset.

The ADC does not transmit $I_i$ and $Q_i$ directly at full rate. Instead, it filters the streams using an FIR filter with a moderately high bandwidth (see the datasheet\cite{adc32rf45ds:2020} for details), then decimates by a factor $D$.  The decimation step transmits one sample every $D$ samples and discards the samples in between.  This step reduces the data rate by a factor of $D$.  In our notation, we replace the sampling frequency $f_s$ by $f_s/D$ for the convenience that in this way, the factor $D$ does not appear in any of the formulas.

Each $\{I,Q\}$ pair received by FPGA forms a complex number $y_i = I_i + \mathbf{j}Q_i$ in the intermediate frequency (IF).  These complex numbers are the recorded raw data in the server, which contain two frequencies $f_0+f_n$ and $f_0-f_n$.  In software, we first apply a filter to remove one of the frequencies, then implement yet another DDC by multiplying $y_i$ with a complex oscillator $\exp\left[\mathbf{j}2\pi(-\frac{p_0}{q_0}\pm\frac{p_1}{q_1})i\right]$.  By selecting an appropriate sign in the equation, this shifts the remaining frequency component to exactly zero (DC).  The exactness is guaranteed by the phase-locking between the signal and the sampling clock.

After this transformation, the output is a stream of complex numbers of nearly time-invariant real and imaginary parts.  We then multiply the each complex sample with a constant $\exp(\mathbf{j}\theta)$ such that the imaginary parts of the stream are minimized towards zero.  The signal amplitude is then calculated through averaging the real parts of the stream in each integration window.

This procedure can also be applied to direct samples $x_i$.  The difference is that the software DDC shall be a real numbered oscillator with the frequency $f_0$.

Through extensive testing with real ADC data, we discovered that the two data processing methods produce very similar results.  RMS calculation is very straightforward but has difficulties in handling arbitrary integration window (see the aforementioned caveats).  Down-convert to DC then average, on the other hand, is more involved but allows arbitrary integration windows.  We use and validate both methods in the subsequent analysis.

\section{RF receiver characterization}
\label{sec:benchtest}

\begin{figure}[!htb]
  \centering
  \begin{minipage}[t]{\linewidth}
    \centering
    \includegraphics{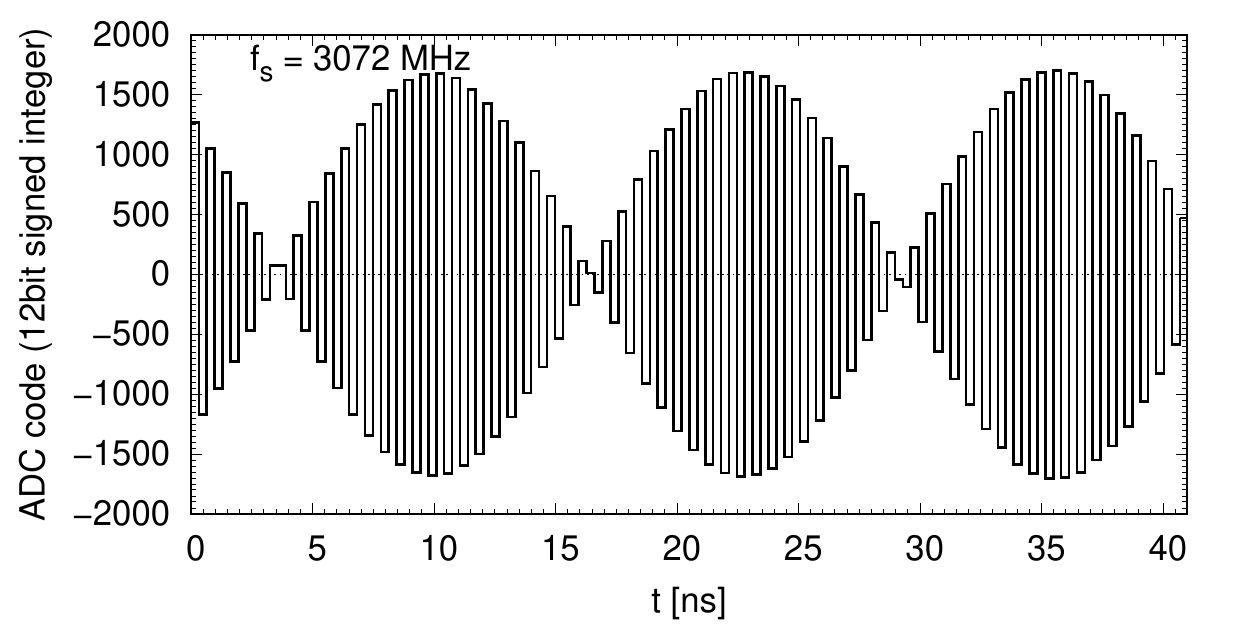}
  \end{minipage}\\
  \begin{minipage}[t]{\linewidth}
    \centering
    \includegraphics{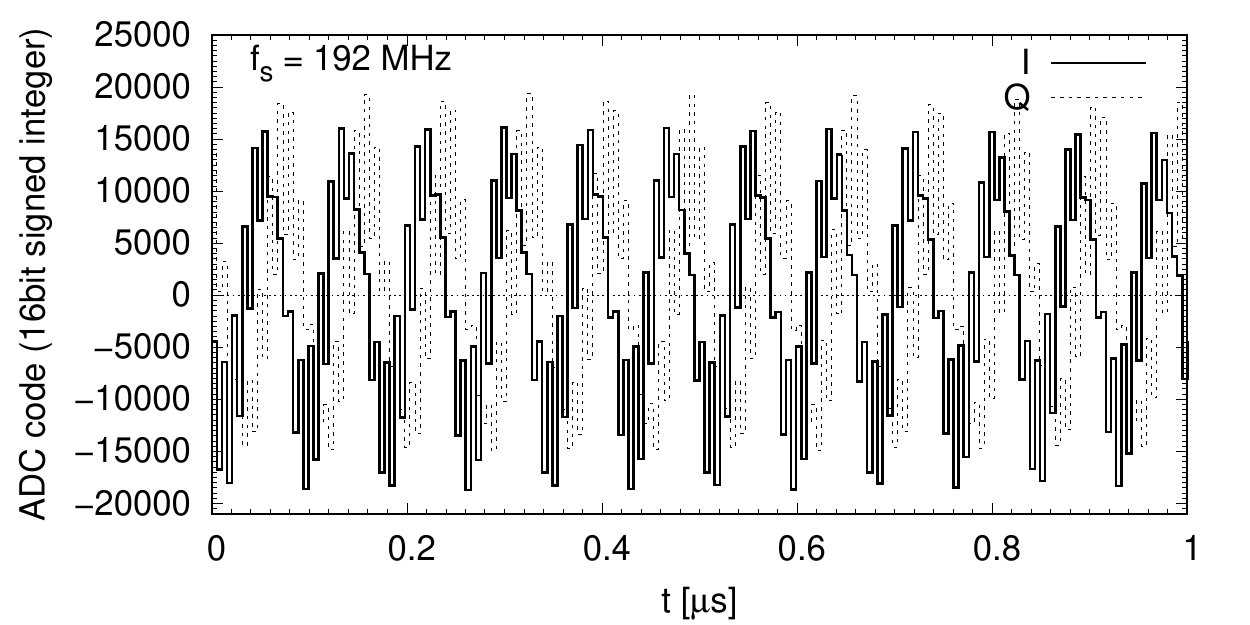}
  \end{minipage}\hspace{0em}%
  \caption{Top: raw waveform of a signal generator produced \SI{1497}{MHz} input signal sampled at \SI{3072}{Msps}.  The signal and the sampling clock are phase-locked via a common \SI{10}{MHz} reference.  Bottom: waveform in DDC/decimation mode.  The signal comes from a BCM cavity, induced by a \SI{130}{\micro A} beam.  The ADC still samples at \SI{3072}{Msps}.  The \SI{1497}{MHz} input signal is digitally mixed with a \SI{1485}{MHz} NCO and decimated by a factor of 16.  $I$ and $Q$ are digitally computed in the ADC and transmitted as 16-bit signed integers.  The effective sampling rate is \SI{192}{MHz}.}
  \label{fig:rawwav}
\end{figure}

We characterize the RF receiver by injecting known signals from signal generators.  The sampling frequency $f_s$ is set to \SI{3072}{MHz} and the injected signal frequency $f_0$ is \SI{1497}{MHz}.  When the signal generator is phase-locked to the sampling clock generator, it guarantees the relation $f_0 = (499/1024)f_s$.  For characterizing the amplitude measurement uncertainty, the signal generator output is split and connected to two ADC channels via a passive RF power splitter.  Correlations between these two channels are used to measure the uncertainty.

A waveform of the raw ADC samples is shown in Fig.~\ref{fig:rawwav} (top).  Since the sampling rate is above \SI{2.5}{Gsps}, only 12 bits per sample are transmitted.  The apparent beating effect is due to the fact that the signal frequency is very close to the Nyquist frequency.  There is only one frequency component in the input signal.

In DDC/decimation mode, the output signal has both $I$ and $Q$ components and two frequencies (mixing products).  The effective sampling rate $f_s$ is reduced by the decimation factor $D$.  A typical waveform in the case $D=16$ is shown in Fig.~\ref{fig:rawwav} (bottom).

In both cases, we record all samples in the computing server, then calculate the amplitudes in integration windows and evaluate the uncertainties offline.

\subsection{Linearity}
\label{sec:bench:lin}

\begin{figure}[!htb]
  \centering
  \begin{minipage}[t]{\linewidth}
    \centering
    \includegraphics{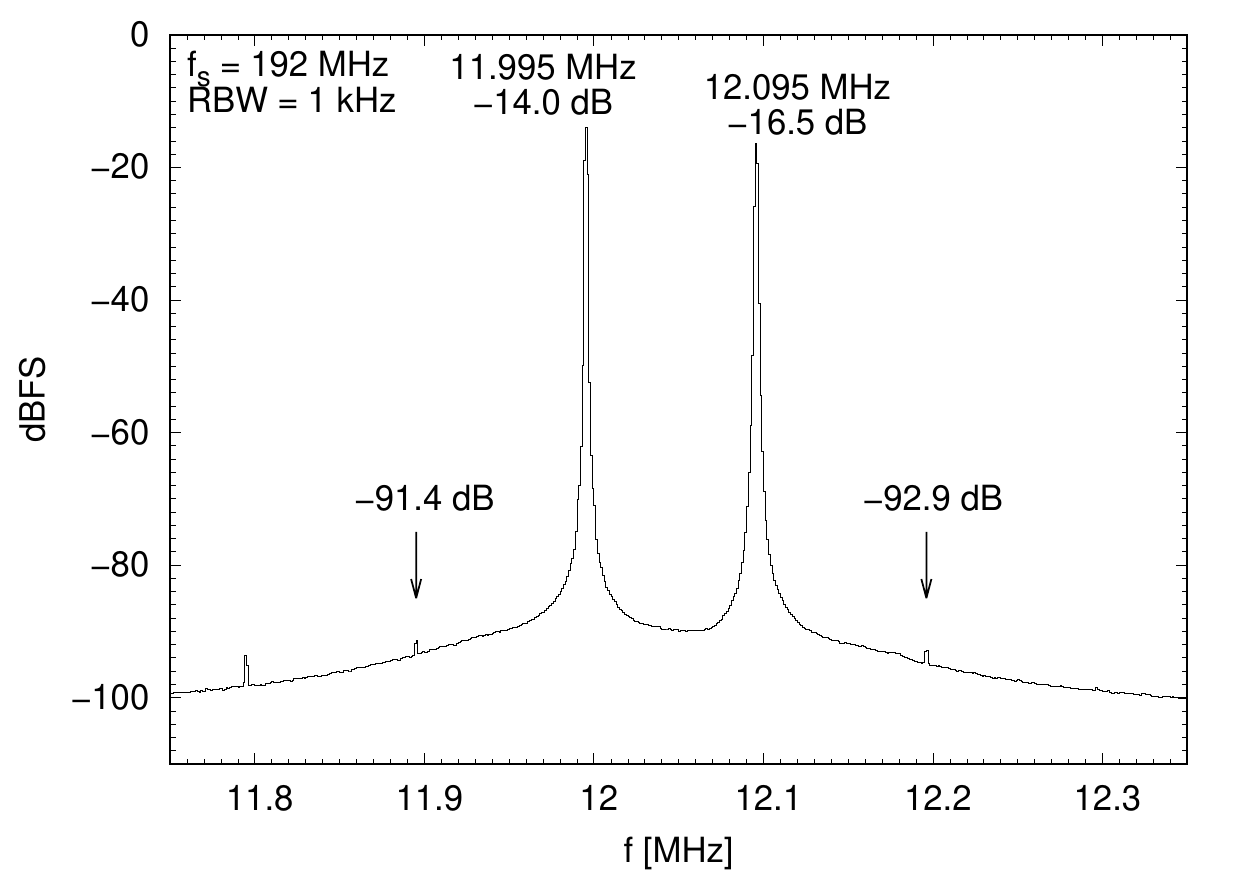}
  \end{minipage}\hspace{0em}%
  \caption{Two tones, 1497.0 and \SI{1497.1}{MHz}, were combined and injected into the receiver for linearity measurement.  The two arrows indicate the locations where the inter-modulation products $(2f_2-f_1)$ and $(2f_1-f_2)$ would appear.  DDC/decimation mode is used therefore the center frequency is shifted to around \SI{12}{MHz} and the effective sampling rate is \SI{192}{Msps}.}
  \label{fig:twotone}
\end{figure}

We employ a two-tone test to characterize the overall linearity of the receiver.  Two signals that are close to each other in frequency are combined and injected into a receiver channel.  We chose the two frequencies to be $f_1=\SI{1497.0}{MHz}$ and $f_2=\SI{1497.1}{MHz}$.  DDC/decimation mode is used to capture the data therefore the center frequency is shifted to around \SI{12}{MHz} and the effective sampling rate is \SI{192}{Msps}.  The amplitudes of the two tones are around $\SI{-15}{dBFS}$.  We measure the intermodulation products resulting from non-linearities.  The results are shown in Fig.~\ref{fig:twotone}.  The intermodulation product amplitudes are around \SI{-92}{dBFS}, which is slightly worse than what is stated in the ADC datasheet\cite{adc32rf45ds:2020}.  The tone and the intermodulation product magnitudes can be combined into the third-order intercept point (IP3) value of about \SI{23.5}{dB}, which translates to the maximum deviation due to non-linearity (3rd order term) to be below $10^{-5}$ of the full-scale magnitude.


\subsection{Amplitude uncertainty}

\begin{figure}[!htb]
  \centering
  \begin{minipage}[t]{\linewidth}
    \centering
    \includegraphics{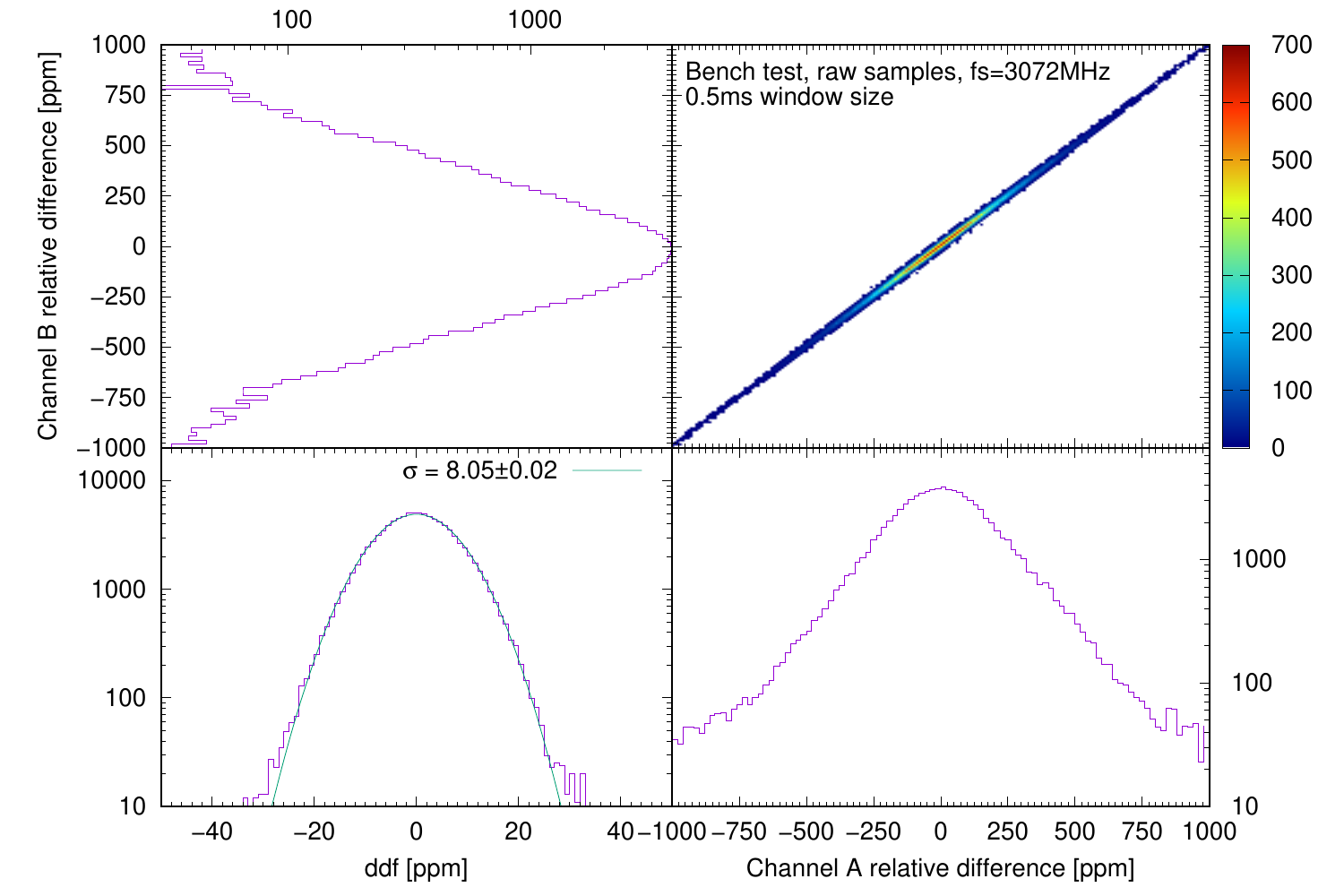}
  \end{minipage}\hspace{0em}%
  \caption{The distribution of relative difference, $rdf$, of each channel, and their correlation double-difference ($ddf$).  The inputs are generated by a signal generator then split passively and fed into each channel.}
  \label{fig:ddfhist2d}
\end{figure}

To assess the amplitude measurement uncertainty, we first compute the signal amplitude in channels A and B that receive the signals split from the same generator.  We denote the amplitudes in each channel as $a_k$ and $b_k$, where $k$ is an integer index labeling the sequence of integration windows.  We define a quantity ``double-difference'' that is the difference between the two channels each measuring the relative difference between consecutive integration windows in their respective channels.  The double-difference can be expressed as
\begin{equation}
    ddf_k = \frac{1}{\sqrt{2}}\left(\frac{a_{k+1}-a_{k}}{a_{k+1}+a_{k}}
           -\frac{b_{k+1}-b_{k}}{b_{k+1}+b_{k}}\right)
\end{equation}
while the relative difference in each channel is expressed as
\begin{equation}
    rdf_k = \frac{a_{k+1}-a_{k}}{a_{k+1}+a_{k}}
\end{equation}

We treat the quantities computed at each $k$ as an individual event and measure the width of the statistical distribution of them.  As shown in Fig.~\ref{fig:ddfhist2d}, the distributions of $rdf$ of both channels A and B, and their correlations, are plotted.  The width of $rdf$ of each channel measures the amplitude fluctuation of the signal generator, which is large and not a good indicator of the RF receiver uncertainty.  However, since the generator signal is split passively and fed into the two channels, by construction, the $rdf$ of the two channels must be highly correlated.  The 2D histogram shown in the top-right panel of Fig.~\ref{fig:ddfhist2d}, which is the correlation between $rdf^{A}$ and $rdf^{B}$, confirms this.  The width of the anti-correlation between the two would largely remove the fluctuations in the original signal and is a good measurement of the receiver uncertainty itself.  The double-difference $ddf$ is exactly constructed to capture this quantity.  The extra factor $1/\sqrt{2}$ is multiplied to normalize the width for the uncertainty contribution from a single channel.

Since the RF amplitude is proportional to the total amount of electrons in the integration window, the quantity $rdf$ directly maps to the asymmetry $A_{PV}$ in MOLLER, and $ddf$ equals the measurement uncertainty in the amount of electrons on an event-by-event (or window-by-window) basis.

\begin{figure}[!htb]
  \centering
  \begin{minipage}[t]{\linewidth}
    \centering
    \includegraphics{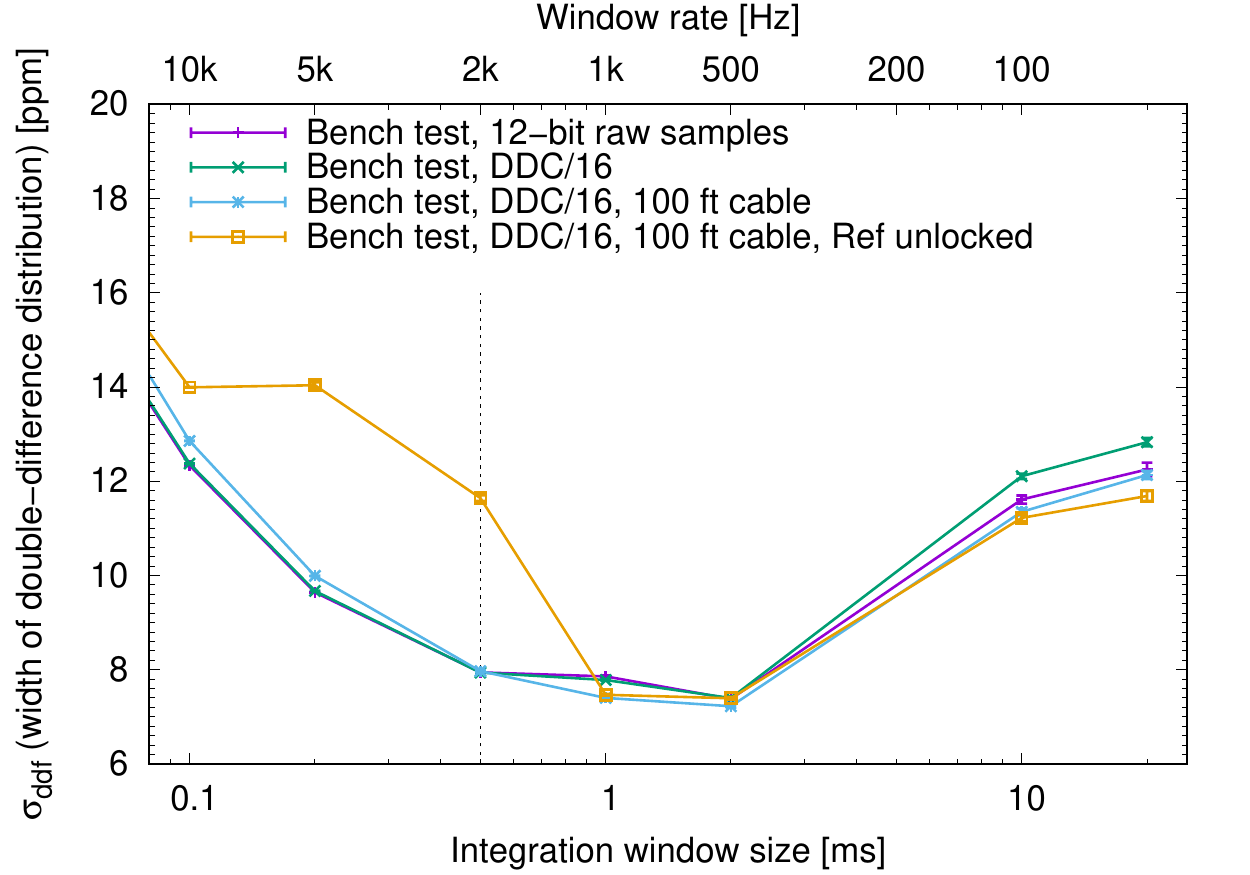}
  \end{minipage}\hspace{0em}%
  \caption{The width of double-difference as a function of integration window size, measured during bench test.  The vertical dashed line indicates \SI{0.5}{ms} integration window, which is the MOLLER requirement.}
  \label{fig:ddfwlen_bench}
\end{figure}

Using signal generator on the bench, we tested both the 12-bit raw data streaming and the DDC/decimation modes as well as software direct RMS calculation and down-convert to DC then averaging.  We measure the width of $ddf$ distribution while varying the integration window size.  The results are shown in Fig.~\ref{fig:ddfwlen_bench}.  A general feature shared among all curves is that the width of $ddf$ is at minimum at an integration window of around \SI{1}{ms}.  It increases as the integration window becomes smaller due to the reduction of averaging (suppression of white noise), and it increases as the integration window becomes larger due to the $1/f$ noise.  This feature is consistent with the expected noise in the system.  Beyond this commonality, other important results from individual tests are summarised below.

First of all, we observe that there is no appreciable difference between direct RMS calculation and down-convert to DC then averaging, therefore the two methods can be used interchangeably when appropriate.  Secondly, the difference between collecting 12-bit raw samples and DDC decimated by a factor of 16 is very small.  Since the DDC/16 mode produces data at a significantly lower rate, it is preferred for the subsequent data taking.

We also tested a configuration that has two 100-ft long LMR-400 coax cables between the two power splitter outputs and each channel.  This is to mimic the conditions at CEBAF where long coax cables of the same type are used to transmit RF signals to the receiver.  The test results show that the long cables do not contribute significantly to the $ddf$ width.

Lastly, we tested the effect of phase lock between signal and sampling clock by intentionally disconnecting the \SI{10}{MHz} reference between the two generators.  Since the phase relation is no longer maintained, in order to eliminate potential window effects in RMS calculations, we use the down-convert to DC then averaging method for analysis.  With the phases unlocked, we observe a substantial increase in $ddf$ width at integration window sizes below \SI{1}{ms}.

\section{Beam test at Jefferson Laboratory}
\label{sec:beamtest}

In September, 2020, we installed the RF receiver in the counting house of Hall A at Jefferson Laboratory and recorded actual beam-induced RF signals during the CREX experiment\cite{e1212004}. To minimize the interference with the running experiment, we took signals from two TM$_{010}$ mode cavities which belong to the beam position monitor (BPM) triplet systems\cite{1nA-BPM} instead of using the standard Jefferson Laboratory beam current monitors (BCMs). The CREX beam has a helicity flip rate of \SI{120}{Hz}.  For the purposes of our tests, the recorded data were not synced to the helicity signal.  Instead, the integration window size is derived from the local clock in the later analysis.

The accelerator is locked to a \SI{10}{MHz} reference clock, which was distributed to our receiver.  The beam and the cavity output are phase-coherent to the reference clock.  We locked the receiver's sampling clock to the same reference lock to achieve coherent sampling.

\begin{figure}[!htb]
  \centering
  \begin{minipage}[t]{0.99\linewidth}
    \includegraphics[width=\linewidth]{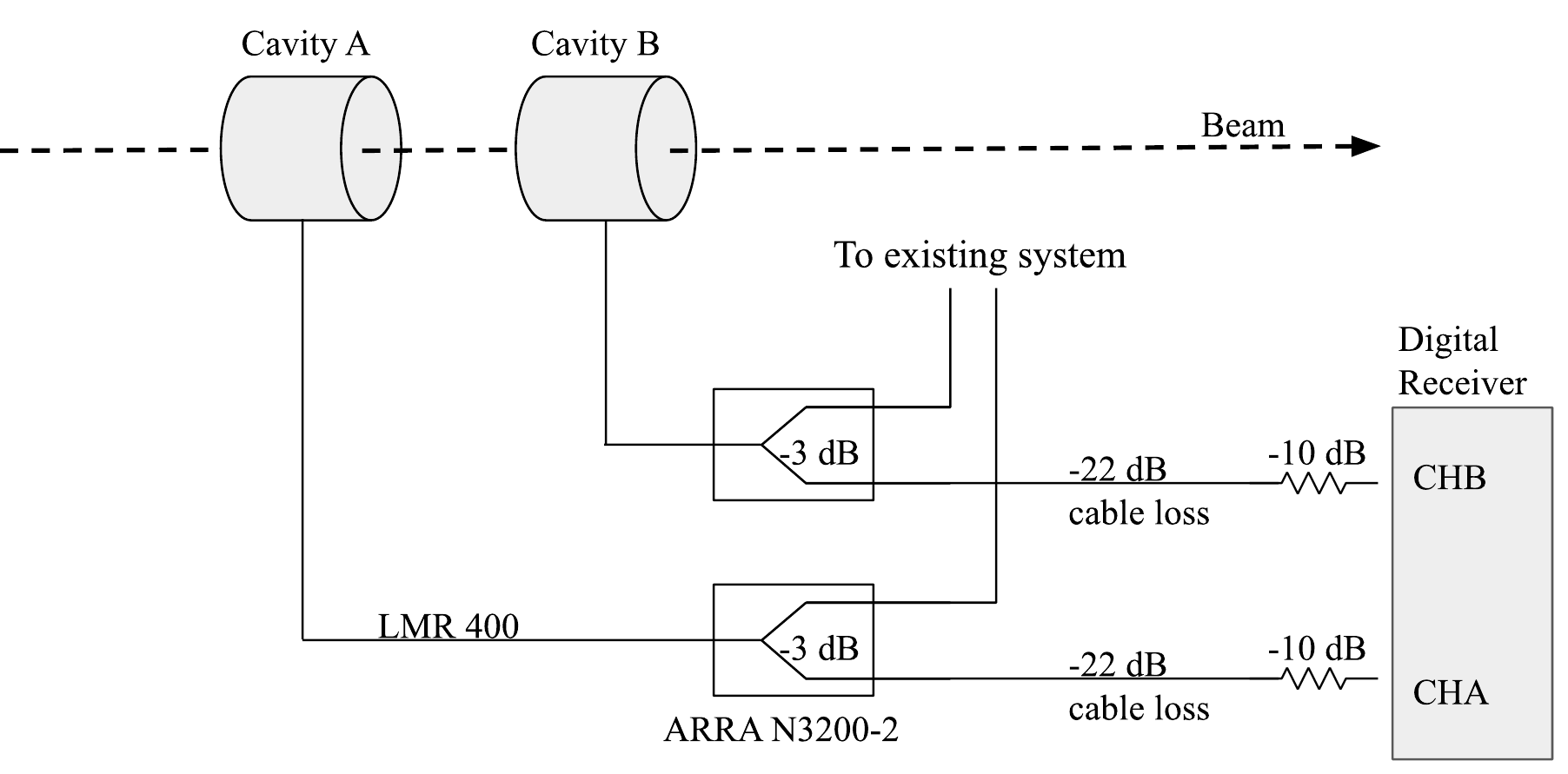}
  \end{minipage}\hspace{0em}%
  \caption{The RF signal chain from the beam cavity to the receiver.}
  \label{fig:Cavity2BCM}
\end{figure}

The relation of the RF receiver to the beam is shown in Fig.~\ref{fig:Cavity2BCM}.  The signals from the two RF resonant cavities installed on the beamline\cite{benesch2021redesigning} are split by a power splitter on each line.  One branch goes to the existing system in the facility; the other branch goes through a long coax cable to the receiver.  The signals incur a power loss of \SI{22}{dB} in the cables.  We opted to install an external \SI{10}{dB} attenuator at each of the inputs for protection and input power level adjustment.

The maximum beam current is about \SI{150}{\micro A}.  The cavity produces an estimated \SI{-2}{dBm} power at the maximum current.  The attenuated signal amplitude after the RF front-end spans about \SI{70}{\percent} of the dynamic range of ADC, which is an acceptable level for reaching a good amplitude uncertainty.  No additional adjustment using the internal step attenuators are necessary.  A typical raw waveform from the beam is shown in Fig.~\ref{fig:rawwav} (bottom).  Data are taken in the DDC/decimate by 16 mode.

\begin{figure}[!htb]
  \centering
  \begin{minipage}[t]{\linewidth}
    \centering
    \includegraphics{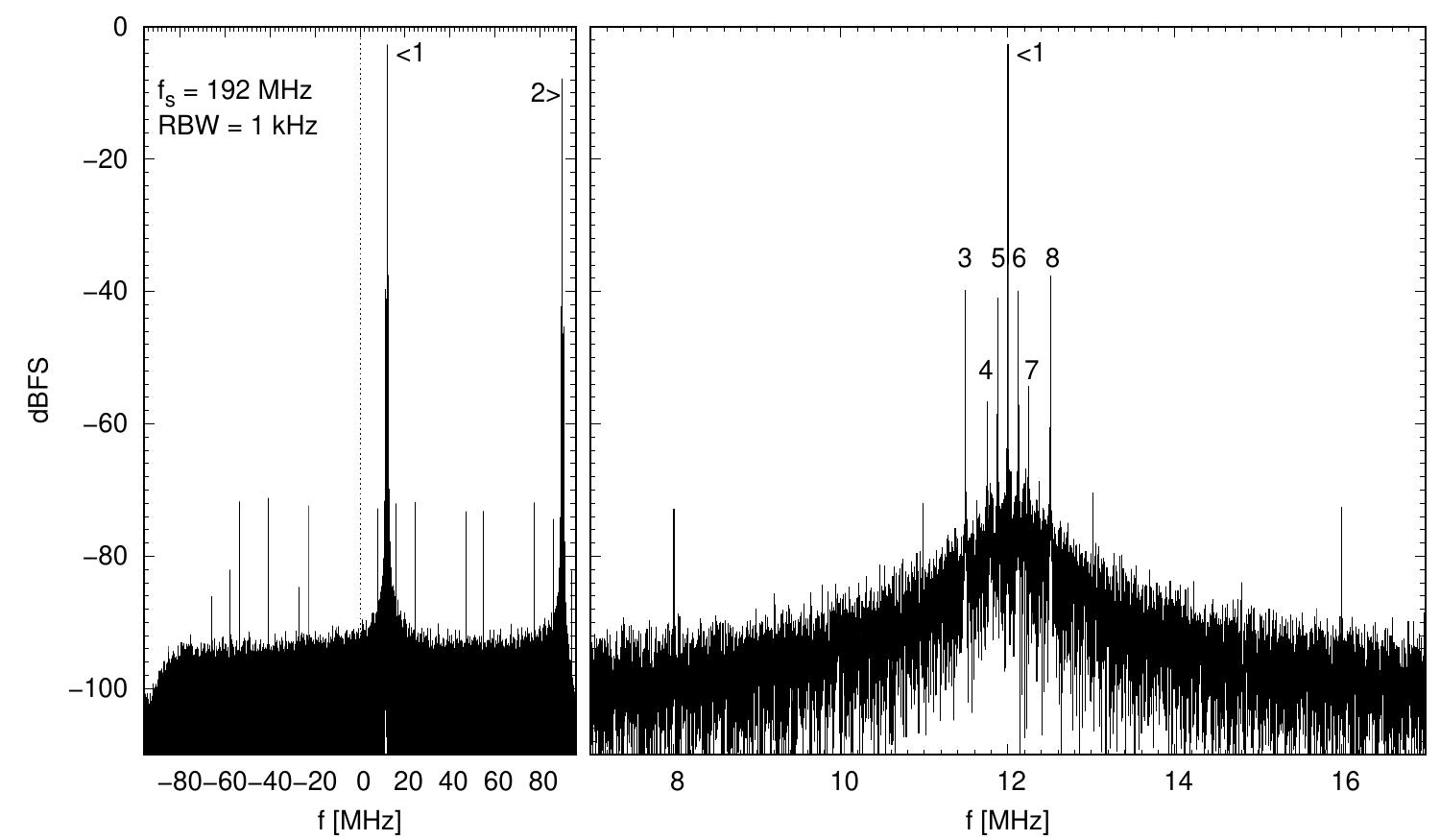}
  \end{minipage}\hspace{0em}%
  \caption{Power spectrum of a representative cavity signal induced by \SI{130}{\micro A} beam.  NCO is set to \SI{1485}{MHz} and is digitally mixed with the input, then decimated by a factor of 16, resulting in an effective sampling rate of \SI{192}{MHz}.  The \SI{1497}{MHz} input signal appears as two mixing product peaks marked 1 and 2 at \SI{12}{MHz} and \SI{90}{MHz} respectively.  Both $I$ and $Q$ are recorded and combined to compute the power spectrum via complex-input FFT, as shown in the left panel.  The right panel shows a zoomed-in view around peak 1, which is of primary interest.}
  \label{fig:psp}
\end{figure}

The power spectrum of a representative signal is shown in Fig.~\ref{fig:psp}.  Due to the NCO and DDC in the ADC, the \SI{1497}{MHz} input signal appears as two peaks at \SI{12}{MHz} (marker 1) and \SI{90}{MHz} (marker 2).  The magnitude of the peaks at markers 3--8 is proportional to that of the main signal and all of these peaks disappear when the beam is off.  While the mechanism of their generation is unknown, these peaks appear to originate from the beam and/or the cavity.
The rest of the spikes are from the ADC and receiver system itself. 

\begin{figure}[!htb]
  \centering
  \begin{minipage}[t]{\linewidth}
    \centering
    \includegraphics{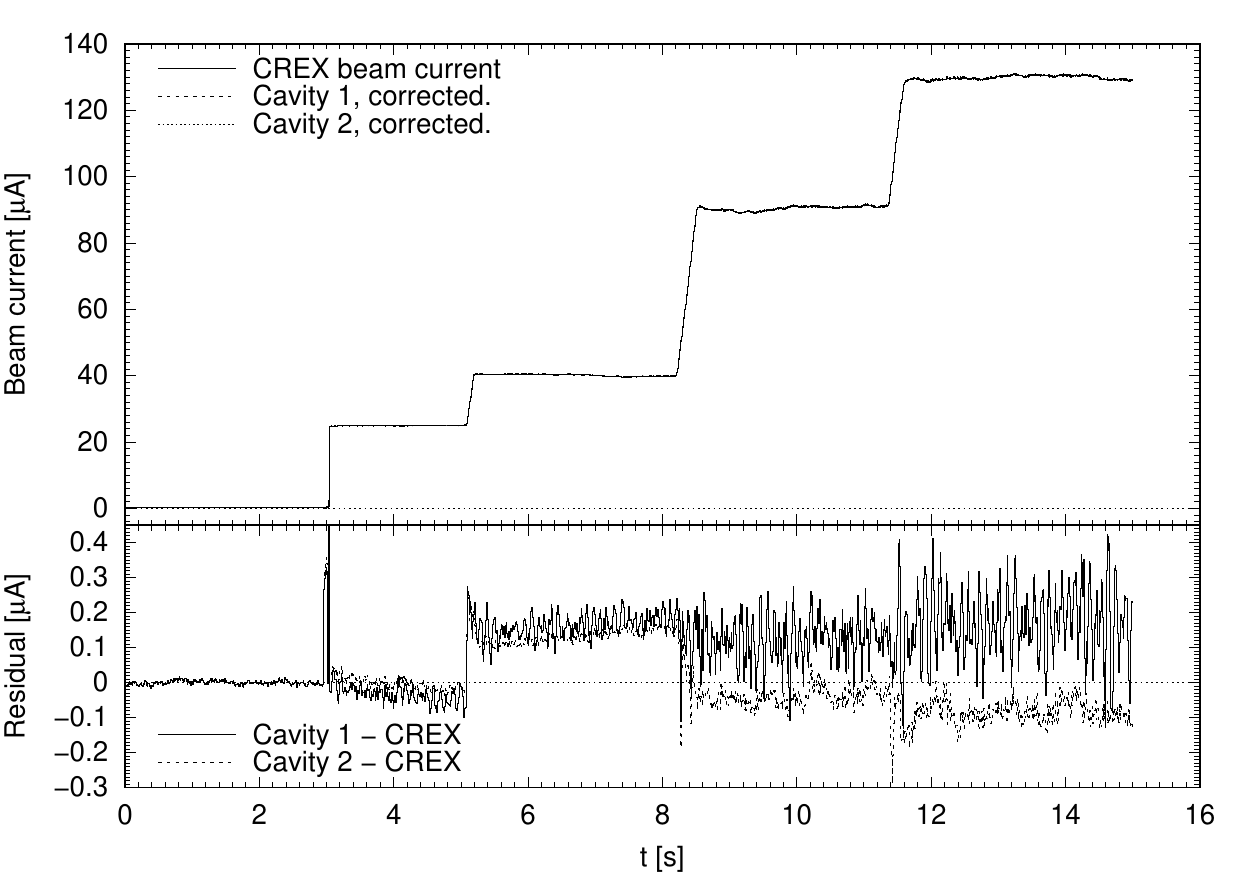}
  \end{minipage}\hspace{0em}%
  \caption{Comparison between beam current measurement provided by the CREX\cite{e1212004} experiment and that inferred from the cavity RF signals measured by this receiver.}
  \label{fig:crexBCMcomp}
\end{figure}

We took data under various beam current levels.  
The CREX experiment provides its own beam current measurement via two BCM cavities.  The BCMs are calibrated against a parametric current transformer (PCT) to reach $1\%$ absolute accuracy\cite{HALLA-BCM}.  We compare the beam current measured by our RF receiver with CREX BCMs.  The results are shown in Fig.~\ref{fig:crexBCMcomp}. The RF receiver measures the amplitudes of RF from both cavities.  Since each cavity and RF channel has slightly different gain, the conversion from RF amplitude to absolute beam current must be tuned independently for every channel.  We assume a linear relation between the RF amplitude and the CREX beam current, and perform a fit in each channel to determine the proportional gain and offset.  The fit parameters are then used to map RF amplitude to the beam current.

The top panel in Fig.~\ref{fig:crexBCMcomp} shows the CREX beam current and the inferred current from RF amplitudes from both cavities.  The three curves are on top of each other and visually indistinguishable.  The differences between inferred currents and CREX beam current are plotted in the bottom panel.  The deviations from a perfect linear relation is generally within \SI{0.2}{\percent}.  This residual is dominated by the quoted linearity of the CREX beam current measurement\cite{kent_private_comm}.  The RF receiver linearity is measured independently to be below $10^{-5}$ (see Sec.~\ref{sec:bench:lin}).  Nevertheless, the RF receiver amplitudes are tracking the beam current well during a beam ramp-up phase as shown in Fig.~\ref{fig:crexBCMcomp}.

\begin{figure}[!htb]
  \centering
  \begin{minipage}[t]{\linewidth}
    \centering
    \includegraphics{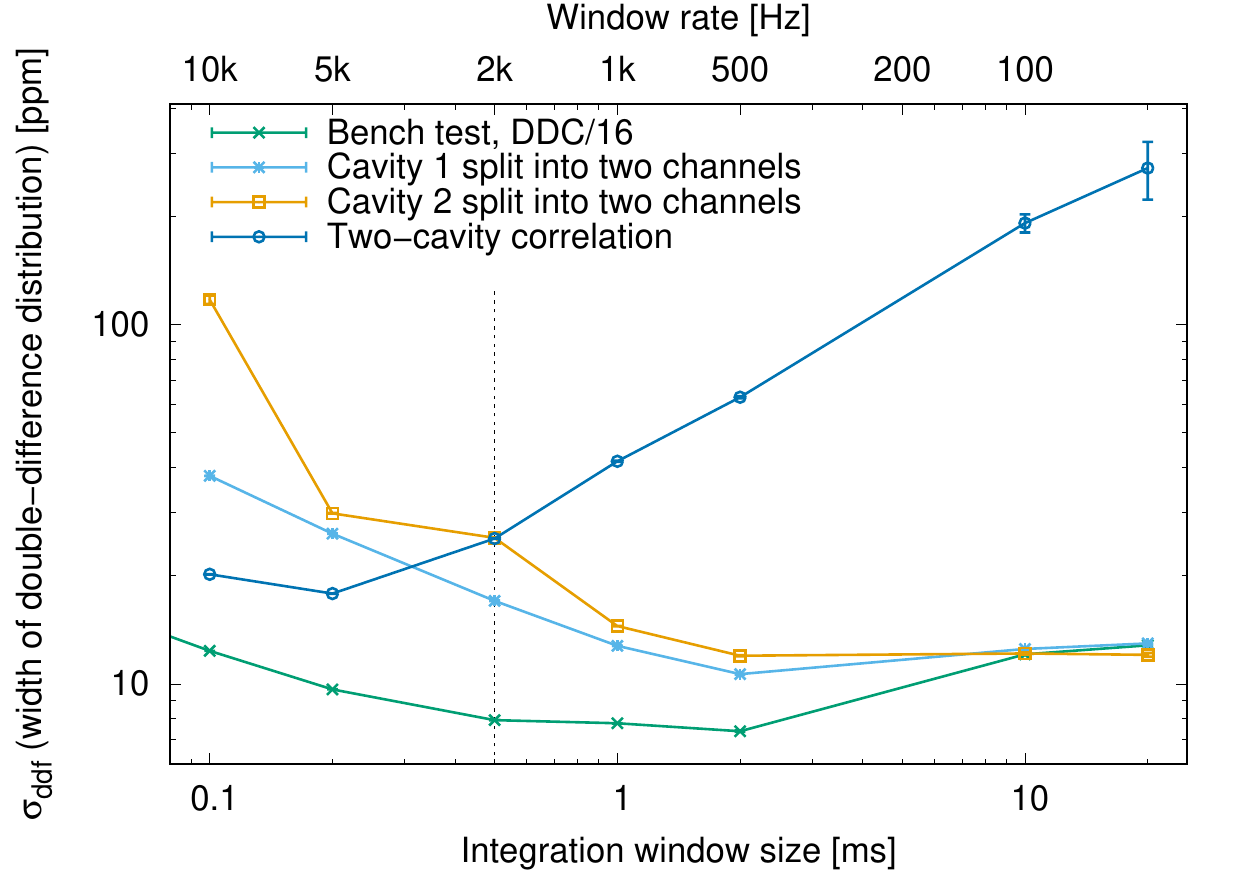}
  \end{minipage}\hspace{0em}%
  \caption{The width of double-difference as a function of integration window size, measured during beam test.  The green curve, labeled ``Bench test, DDC/16'', is identical to the one in Fig.~\ref{fig:ddfwlen_bench} for reference.  The vertical dashed line indicates \SI{0.5}{ms} integration window, which is the MOLLER requirement.}
  \label{fig:ddfwlen}
\end{figure}

The double-difference between two cavities would directly measure the uncertainty in beam charge asymmetry.  We also repeated some of the receiver characterization measurements described in Sec.~\ref{sec:benchtest} using cavity signals in place of signal generators, and split one cavity signal to feed two receiver channels.  This is a useful test to verify if the receiver is impacted by environmental interference in the facility.  The data processing procedure is described in Sec.~\ref{sec:dp}.

The double-difference versus integration window is shown in Fig.~\ref{fig:ddfwlen}.

\begin{figure}[!htb]
  \centering
  \begin{minipage}[t]{\linewidth}
    \centering
    \includegraphics{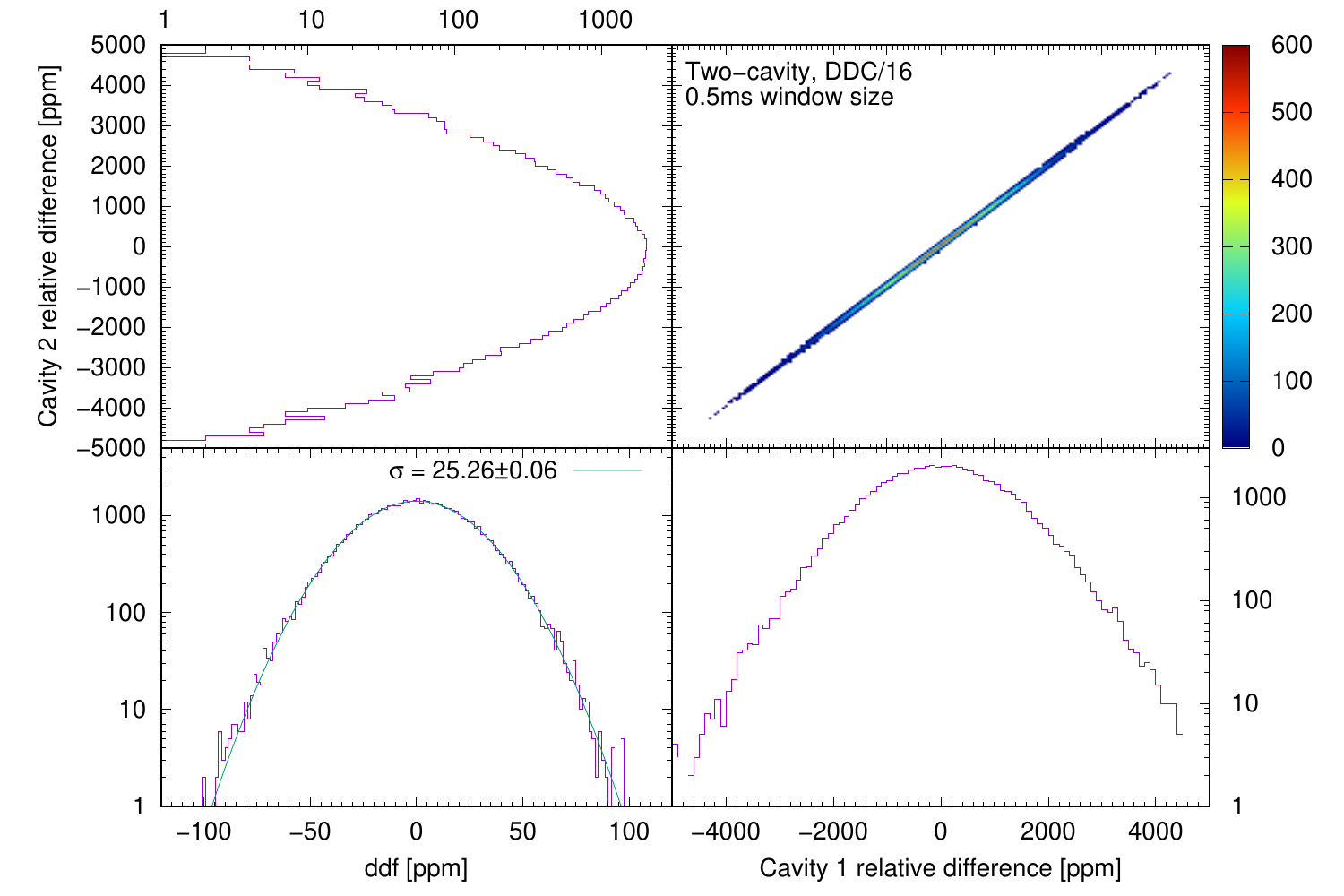}
  \end{minipage}\hspace{0em}%
  \caption{The distribution of relative difference, $rdf$, of each cavity at \SI{130}{\micro A} beam current, and their correlation double-difference ($ddf$).  The integration window size is \SI{0.5}{ms}.}
  \label{fig:ddfhist2d_beam}
\end{figure}

\section{Conclusions and Outlook}

We have presented a direct-sampling RF receiver with an internal noise performance that satisfies 
the stringent requirements of the \MOLLER experiment.  With the integration window of \SI{0.5}{ms} (helicity-flip frequency of \SI{2}{kHz}), we demonstrate an RMS noise of the pulse-pair charge asymmetry of \SI{8}{ppm} per channel, which corresponds to an uncertainty on the measurement of the beam charge of $\sigma(Q)/Q=\SI{11}{ppm}$ in \SI{2}{kHz} bandwidth.  This is the lowest-noise measurement of beam charge among comparable RF systems.  The noise and linearity of the RF receiver are adequate for applications in parity-violation measurements. 

When comparing the beam-induced RF signals measured by two independent cavities in the Jefferson Laboratory Hall A beamline, we observe excess noise in the distribution of the charge asymmetry double-difference, producing an asymmetry double-difference RMS of \SI{25}{ppm} in the \SI{2}{kHz} bandwidth.  Moreover, we observe an increase in the asymmetry double-difference RMS at smaller bandwidths, indicating noise sources that scale with helicity-flip frequency $f$ as $\sim1/f$.  The origin of these noise sources is unclear, but they are inconsistent with the phase noise in the RF receiver, or diurnal effects induced in the cables.  One potential cause is the variation of the RF signal amplitude in the cavity due to beam position jitter, induced by the variation of the RF field with beam position.  To explain the observed $\mathcal{O}(\SI{20}{ppm})$ excess noise in the charge double-difference, the two RF cavities would have to be misaligned relative to each other by several millimeters, and the beam position jitter would have to grow at low frequencies as $1/f$.  Variations in the beam losses between the two cavities could also contribute to the systematic uncertainty in the measurement of beam charge.  These sources will be further investigated in subsequent beam measurements at Jefferson Laboratory.

\acknowledgments

The design and construction of the RF receiver was supported by the Laboratory Directed Research and Development (LDRD) funding from Berkeley Lab, provided by the Director, Office of Science, of the U.S.\ Department of Energy under Contract No.~DE-AC02-05CH11231.  We thank Caryn Palatchi, Kent Paschke, John Musson, and Jefferson Laboratory staff for the assistance provided during the beam test.  We acknowledge the support and fruitful discussions from LBNL engineering staff Kerri Campbell, Larry Doolittle, and Vamsi Vytla.  Various discussions with the MOLLER collaboration inspired this work.  We thank the CREX collaboration for providing the beam current data.


\clearpage


\bibliographystyle{JHEP}
\bibliography{refs}

\providecommand{\href}[2]{#2}\begingroup\raggedright\begin{thebibliography}{10}

\bibitem{Benesch:2014bas}
{\scshape MOLLER} collaboration, \emph{{The MOLLER Experiment: An Ultra-Precise
  Measurement of the Weak Mixing Angle Using M{\o}ller Scattering}},
  \href{https://arxiv.org/abs/1411.4088}{{\ttfamily 1411.4088}}.

\bibitem{Farkas:1976slac}
Z.D.~Farkas, H.A.~Hogg, H.L.~Martin and A.R.~Wilmunder, \emph{{Recent
  Developments in Microwave Beam-Position Monitors at SLAC}},
  {\emph{SLAC-PUB-1823} (1976) }.

\bibitem{Whittum:1998uc}
D.H.~Whittum and Y.~Kolomensky, \emph{{Analysis of an asymmetric resonant
  cavity as a beam monitor}},
  \href{https://doi.org/10.1063/1.1149756}{\emph{Rev. Sci. Instrum.} {\bfseries
  70} (1999) 2300}.

\bibitem{Qweak:2014xey}
{\scshape Qweak} collaboration, \emph{{The Q$_{weak}$ experimental apparatus}},
  \href{https://doi.org/10.1016/j.nima.2015.01.023}{\emph{Nucl. Instrum. Meth.
  A} {\bfseries 781} (2015) 105}
  [\href{https://arxiv.org/abs/1409.7100}{{\ttfamily 1409.7100}}].

\bibitem{adc32rf45ds:2020}
{\relax Texas Instruments}, \emph{ADC32RF45 datasheet}.
\newblock https://www.ti.com/product/ADC32RF45.

\bibitem{vcu108:2020}
{\relax Xilinx}, \emph{VCU108 datasheet}.
\newblock https://www.xilinx.com/products/boards-and-kits/ek-u1-vcu108-g.html.

\bibitem{erasynth:2020}
\emph{ERASynth datasheet}.
\newblock
  https://github.com/erainstruments/erasynth-docs/blob/master/erasynth-datasheet.pdf.

\bibitem{dpdk}
{Linux Foundation}, \emph{Data Plane Development Kit ({DPDK})}, 2015.
\newblock http://www.dpdk.org.

\bibitem{e1212004}
S.~Covrig~Dusa, J.~Mammei, D.~McNulty, K.~Paschke, S.~Riordan and P.~Souder,
  \emph{{CREX: parity-violating measurement of the Weak Charge distribution of
  $^{48}$Ca to 0.02\,fm accuracy}}, {\emph{Jefferson Lab Experiment Proposal
  E12-12-004} (2013) }.

\bibitem{1nA-BPM}
R.~Ursic, R.~Flood, C.~Piller, E.~Strong and L.~Turlington, \emph{{1 nA beam
  position monitoring system}},  in \emph{Proceedings of the 1997 Particle
  Accelerator Conference (Cat. No.97CH36167)}, vol.~2, pp.~2131--2133 vol.2,
  1997, \href{https://doi.org/10.1109/PAC.1997.751131}{DOI}.

\bibitem{benesch2021redesigning}
J.~Benesch and Y.~Roblin, \emph{{Redesigning the Jefferson Lab Hall A Beam Line
  for High Precision Parity Experiments}},
  \href{https://arxiv.org/abs/2109.05996}{{\ttfamily 2109.05996}}.

\bibitem{HALLA-BCM}
J.-C.~Denard, A.~Saha and G.~Laveissiere, \emph{{High accuracy beam current
  monitor system for CEBAF's experimental Hall A}},  in \emph{PACS2001.
  Proceedings of the 2001 Particle Accelerator Conference (Cat. No.01CH37268)},
  vol.~3, pp.~2326--2328 vol.3, 2001,
  \href{https://doi.org/10.1109/PAC.2001.987367}{DOI}.

\bibitem{kent_private_comm}
K.~Paschke. {Private Communication}, 2021.

\end{thebibliography}\endgroup








\end{document}